\pdfoutput=1 
\documentclass{article}

\usepackage{graphicx}
\usepackage[a4paper,vmargin={20mm,20mm},hmargin={20mm,40mm}]{geometry}
\usepackage[margin=2cm]{caption}

\title{Warp2: A Method of Email and Messaging with Encrypted Addressing and Headers}

\author{H. Bjorgvinsdottir$^a$ \\
 P. M. Bentley$^{a,b}$ \thanks{Corresponding author: phil.m.bentley@gmail.com} \\
\small {$^a$}University of Uppsala, Uppsala, Sweden\\
\small {$^b$}European Spallation Source ESS AB, Box 176, Lund 244 10, Sweden\\
}

\begin{document}

\maketitle

\abstract{Secure communications are playing increasing roles in society, particularly in finance, journalism, and military projects.  Current methods of securing e-mail and similar messaging methods rely on encryption of the message body, but the header with addressing information remains plaintext.  This allows third party eavesdroppers to collect and analyse the header metadata and construct a network model of the participants in conversations (who, where, when, subject).  In this article, we describe a method of communication where the header is also encrypted, hindering the assembly of the communication network models, which is verified with a working prototype application.  This provides a useful tool to journalists and proponents of free speech in oppressed countries, protecting both the messages and their sources.}

\newpage
\section{Introduction}\label{sec:Introduction}
Digital communications are widespread, and the norm for both business and personal interaction in the modern world.  The most widely-used email and messaging systems pass messages entirely or partially in plaintext, allowing eavesdroppers at some stage the opportunity to harvest the message content to their own ends.  Common uses are the collection of contact details for spam lists, analysing message content for targeted advertisements, and occasionally financial fraud or government surveillance.  More recently, the scope of government surveillance in some oppressed political regimes has become the subject of concern for those supporting democracy, free speech, and holding leaders accountable via journalism.
\begin{figure}[hbt]
\centering
\includegraphics[width=0.8\textwidth]{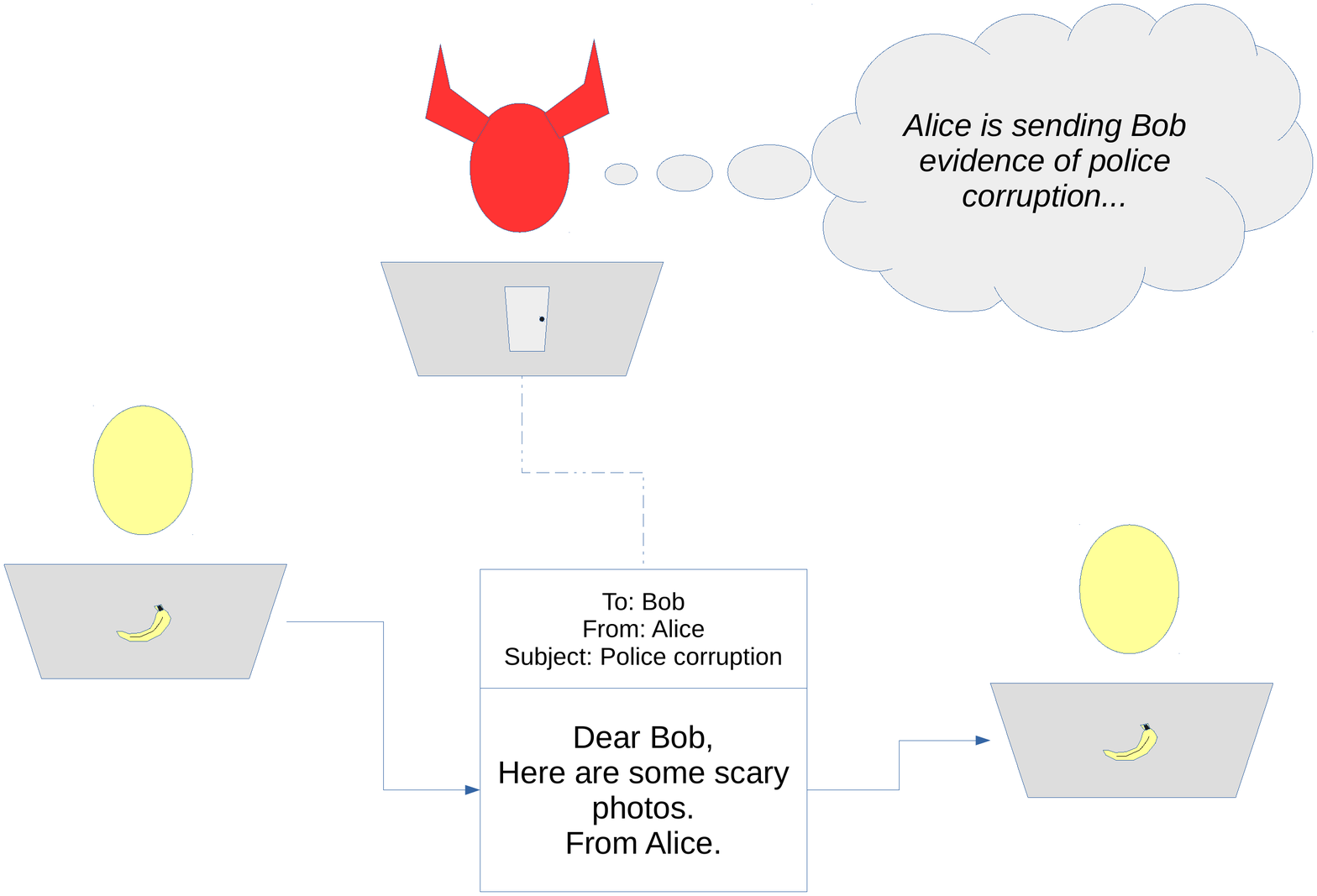}
\caption{Illustration of email without encryption.  An adversary can intercept the message and read it.}
\label{fig:plainEmail}
\end{figure}
\begin{figure}
\centering
\includegraphics[width=0.8\textwidth]{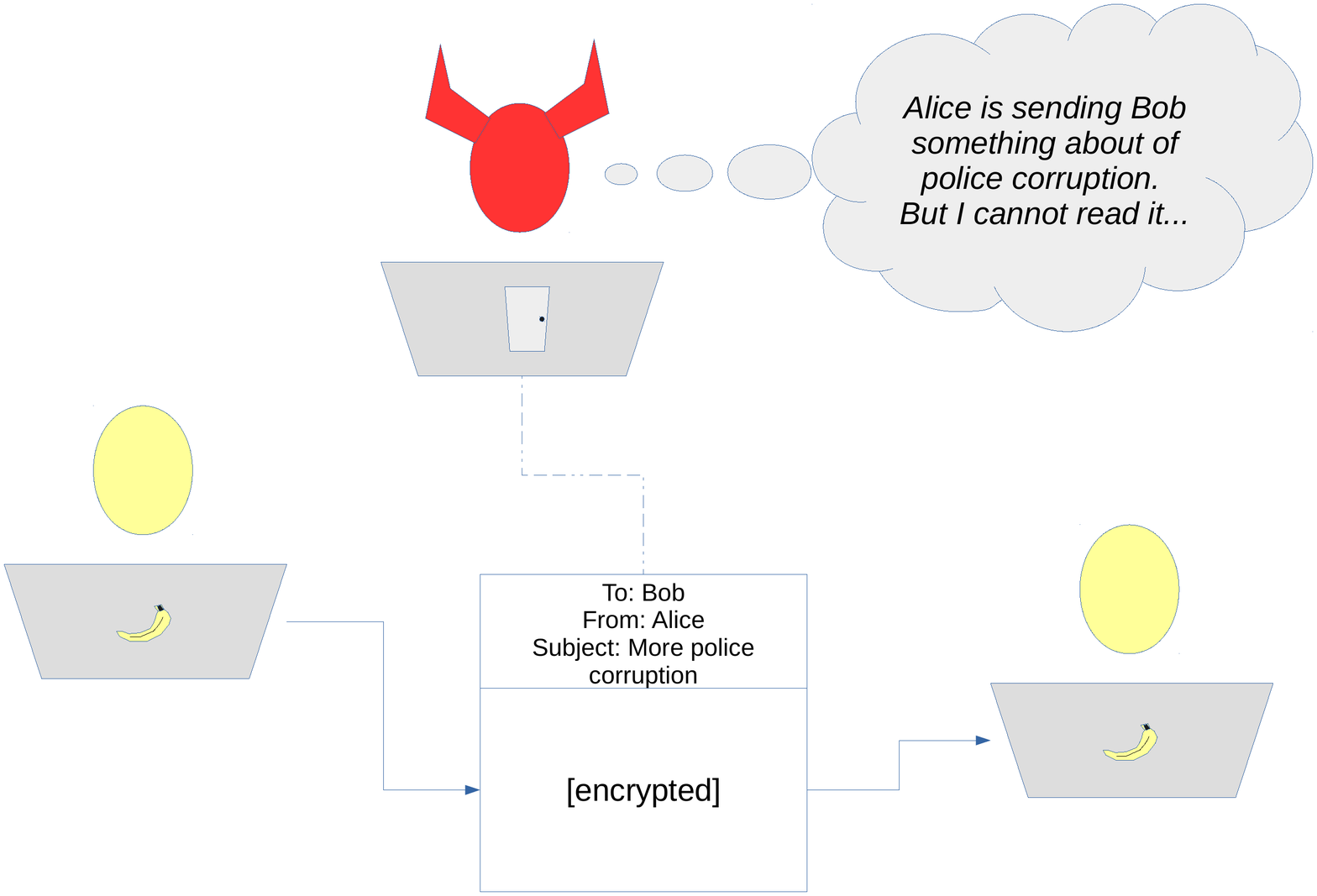}
\caption{Illustration of email with currently available encryption.  An adversary can intercept the message but can only read the header, not the content.}
\label{fig:gpgEmail}
\end{figure}

One solution to this problem is to encrypt the communications using end-to-end public-key encryption, illustrated in figure \ref{fig:gpgEmail}.  

The sender encrypts the message on their device, and only a recipient with the corresponding private key is able to decrypt the message on their own device.  The most famous example is the RSA method \cite{RSA_ENCRYPTION}, which is freely implemented in GPG \cite{GNUPG}.  Using encrypted mail, the message content and any attachments are protected with strong algorithms to prevent third parties from reading the content of the message easily.  However, this still leaves the ``metadata'', which is the message header.  This contains information about the sender, the recipient, the time of the correpsondence, the given subject of the message, and some of the routing that the message took.  Third parties who intercept this metadata are able to reconstruct the spheres of communication in the social network of an individual: who they talk to, when and how frequently, and about what, even if the details of the messages are protected.  The metadata can be gathered at each and every step of the chain of servers that transmit the messages across the internet.  This means that for journalists, for example, who wish to keep the identity of their sources confidential, or for government agents who wish to be discreet about their correspondence, by using email they risk exposing themselves and their message recipients to unwanted attention from third parties.

To protect the users identities and the rest of the communication network, it would be necessary to encrypt the message header with strong encryption.  In this way, it is more difficult to establish who is talking to whom, and the source and recipient of the message can be protected.  The primary problem to solve is the passing of a message when the identity of the recipient is unknown --- where do you send it?  The secondary problem is to make sure the system is sufficiently fast for users.  In this article, we describe a prototype system along these lines, and the limitations of the current method, which may be improved considerably in future work.

\section{Description of Method}
\begin{figure}
\centering
\includegraphics[width=0.8\textwidth]{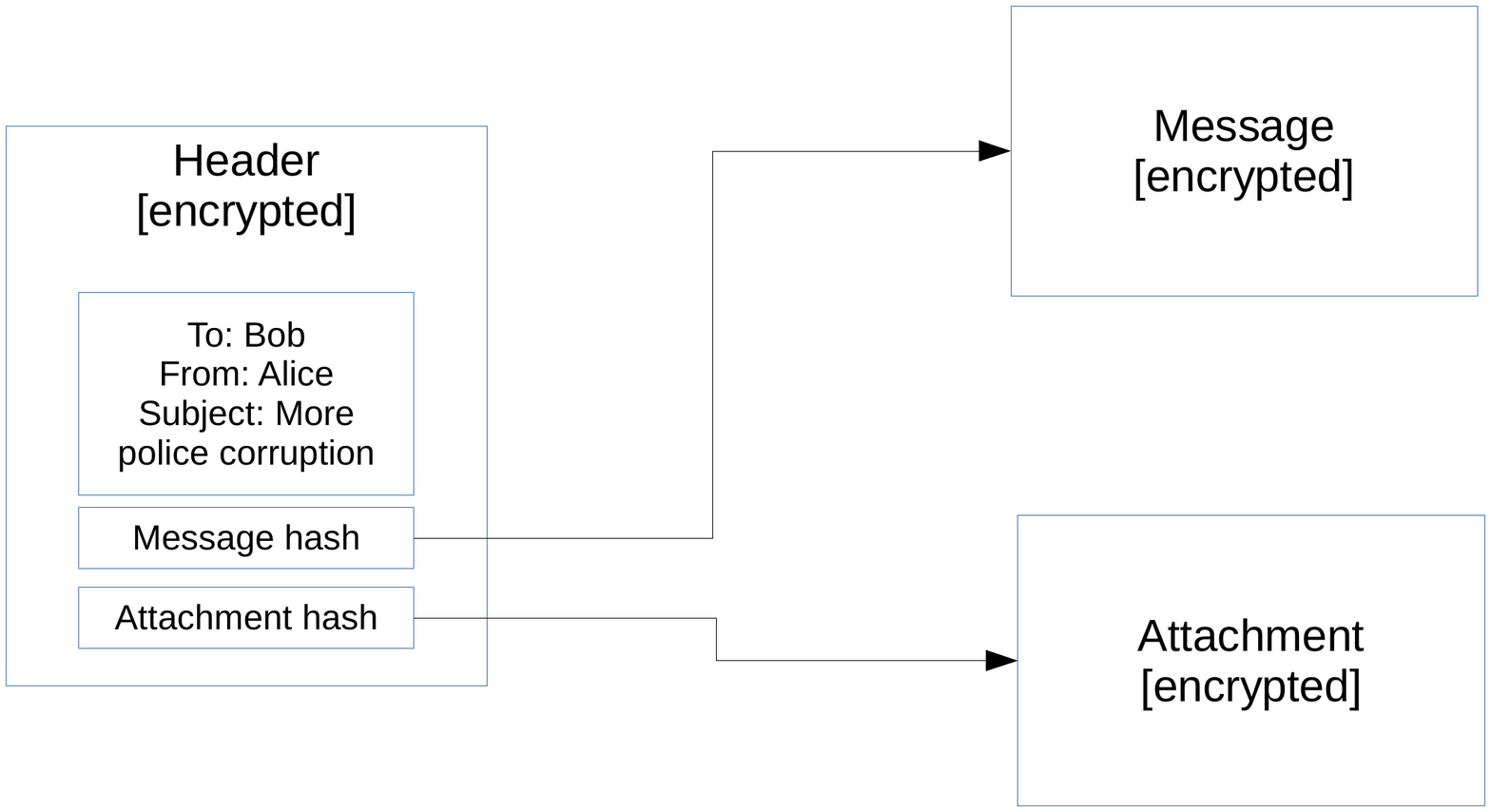}
\caption{Arrangement of files for a warp2 message.}
\label{fig:schematic}
\end{figure}

Our method, which we call ``warp2'' uses GnuPG to encrypt the message and attachments like a regular encrypted email.  The plaintext of the message header contains \emph{to, from, date, subject} fields like a regular email message.  In addition to the usual header data, warp2 headers also contain unique SHA2 identifiers for separate files making up the message body and an attachment.  The arrangement of files is illustrated in figure \ref{fig:schematic}

Traditional messaging systems have separate inboxes for each user.  If that were the case for warp2, monitoring the access to an inbox could associate a user with that address, even though the message and headers are encrypted.  The solution is to have \emph{one} public inbox for all users on the server, containing all current message headers.

Our prototype uses GPG to encrypt both the message contents and the headers.  Each encrypted header file is around 1 kb in size.  All users synchronise a copy of the entire header inbox on their device.  The ``inbox'' is a very simple database of header files, in our case stored with filenames that are SHA2 hashes of the encrypted header file.  In the prototype, a simple MYSQL and PHP implementation was used to demonstrate the server methodology.  The encrypted message body and attachment files are stored in separate fields in the database, and these are not downloaded in full.  That way, the amount of traffic is minimised to just the message headers.

\subsection{Sending a Message}
The device of the sender creates two or three files: the header, the message body, and an attachment if necessary.  The user encrypts the message body and attachment, and calculates a SHA2 hash of the encrypted body and attachment files.  These are put into the message header file.   A SHA2 hash of the message header plaintext is created to allow proof of receipt (see section \ref{sec:speed} later).  The header is then encrypted.  All three files are uploaded to the remote inbox database, along with hash of the plaintext of the header.  

\subsection{Receiving a Message}
The user first synchronises the inbox by downloading all new encrypted message headers from the common inbox over an encrypted connection.  Next, they filter the message headers to identify which messages are intended for them.  Warp2 tries to decrypt every header, and flags any headers that fail decryption to be ignored in future.  On the other hand, successful decryption reveals the plaintext header, containing the identifiers of the message body and attachment.  The recipient then downloads the message body and attachment, and decrypts them.  Finally, the recipient computes the SHA2 hash of the plaintext header, and notifies the server of this plaintext.  The server can then remove the message files associated with this plaintext hash, or flag them as sent.  The message sending process has now completed.

\begin{figure}
\centering
\includegraphics[width=0.8\textwidth]{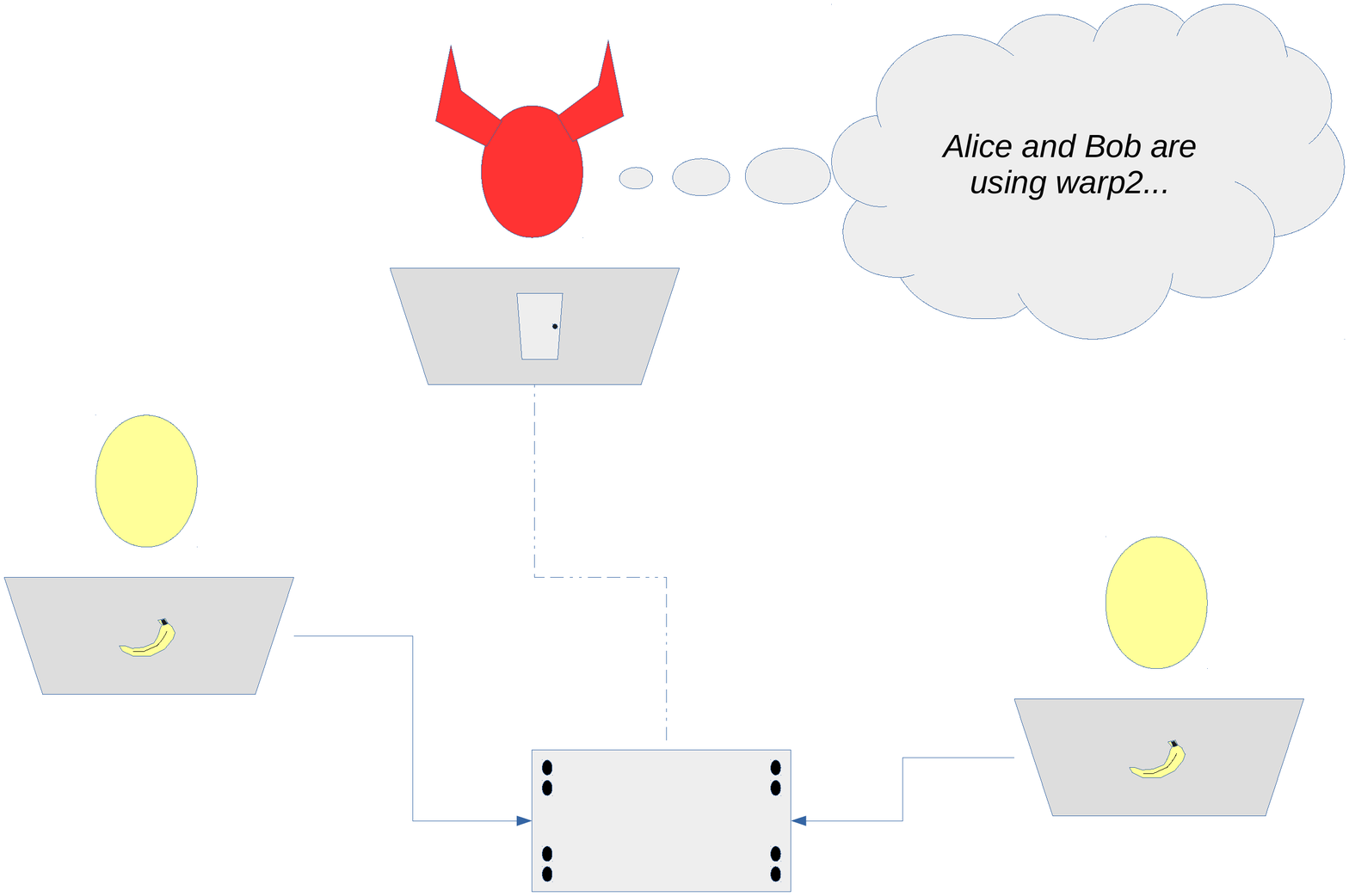}
\caption{An adversary may be able to identify users of a warp2 server by intercepting web traffic, but not glean information about the conversation.}
\label{fig:comms}
\end{figure}

An adversary may be able to establish that one or more individuals are using a warp2 server, but they cannot easily identify the participants of a particular conversation just by examining the messages alone, as shown in figure \ref{fig:comms}.

If the adversary polls the server to download the messages, they are free to do so, but they do not obtain anything more than a large number of separate encrypted files, unless the have other means of breaking the encryption.  
\begin{figure}
\centering
\includegraphics[width=0.8\textwidth]{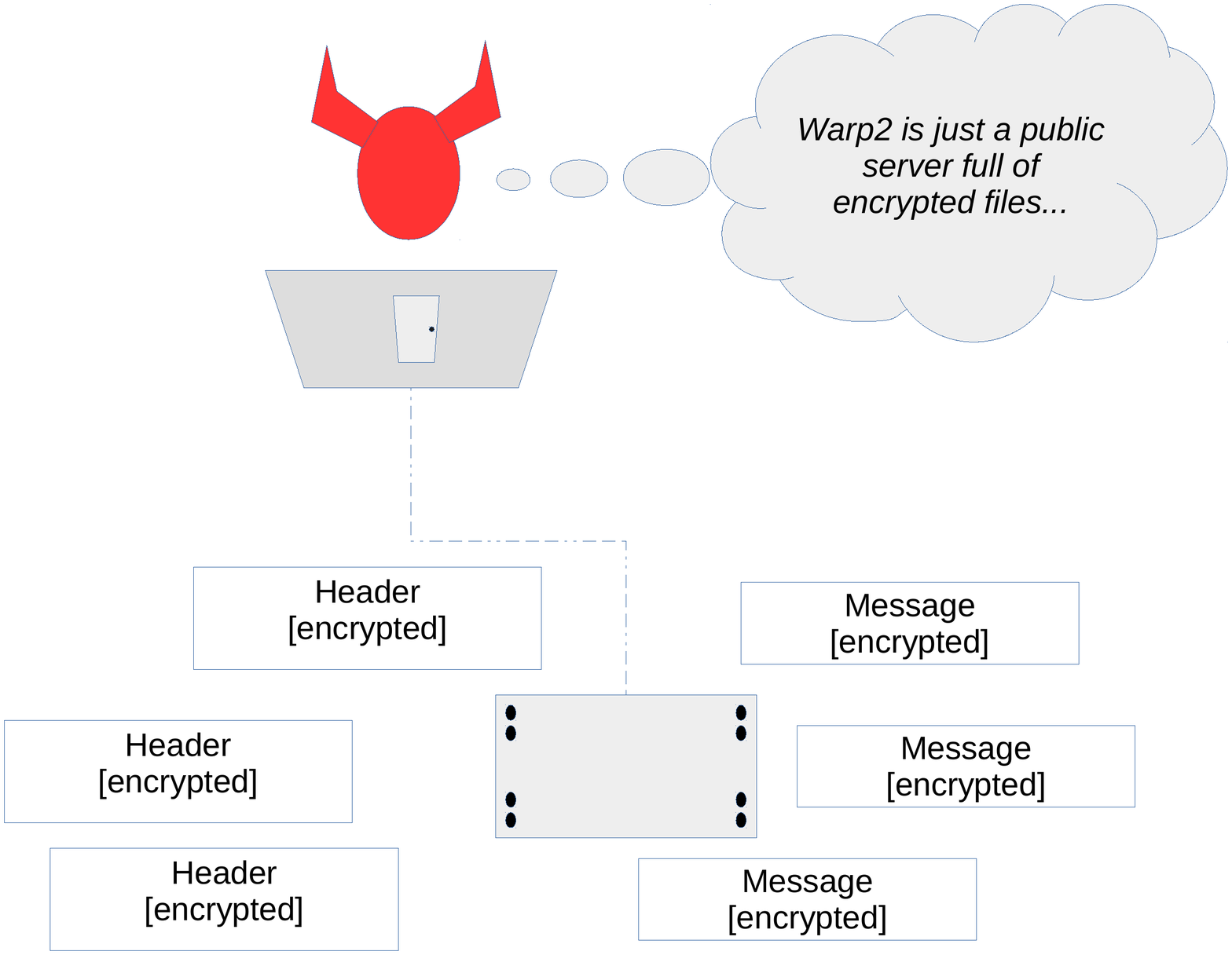}
\caption{Scanning of the server reveals an open server full of encrypted files.}
\label{fig:webserver}
\end{figure}

\section{Discussion}

\subsection{Speed Issues\label{sec:speed}}
The warp2 headers in the prototype using GPG are 1 kb in size.  Assuming 1000 users on the server, sending around 10 messages per day, the client software needs to decrypt less than $\sim$10 MB of header data per day, and the daily server bandwidth would be less than $\sim$10 GB assuming the users check their inbox once per day.

Reducing the number of stored messages to the minimum necessary also helps to keep the server and network requirements down.  One way is to notify the database to purge messages once they have been read by the recipient.  This would also expose the system to vandals who could delete any message they wanted, unless proof of readership is obtained.  The message database contains a hash of the header plaintext (not the plaintext itself).  The idea is that when the recipient calculates their computed hash of the header plaintext, they send it to the database which deletes the row of a matching message, since the message transmission is complete\footnote{This feature is partially implemented in the prototype application, the server side is missing.}.

\subsection{Obvious Potential Attacks}
The method is only as reliable as the weak points in the chain, and warp2 is exposed to the same risks as the encryption technology, in the case of the prototype this is GPG.

Instead of tracking the headers, it is possible for a third party to track the private keys that are needed to decrypt the message as a form of metadata, and construct a social network model around private key IDs rather than email addresses.  To defend against this, a double key exchange can be envisaged.  The first key exchange is made by external means, e.g. email, face to face meeting, just like regular encrypted email.  The second key exchange happens within warp2, where the participants send their new keys to each other in a single message, and delete their old key pairs.  This creates two key pairs that are anonymous to a third party, and independent of the initial point of contact, unless the warp2 messages making the key exchange are compromised by brute force or by duplication of keys.  Re-exchanging keys this way can be done as frequently as required, even per message for the super-cautious.  Analysis of the server traffic then may only allow identification of a fairly large group of IP addresses who are using the server.

An attacker may be able to identify users from analysing the majority of the network traffic to the server, and simultaneously polling the server at regular intervals whilst monitoring target users.  By linking the timing of new messages appearing with increased encrypted traffic of the target user, it may be possible to statistically associate an IP address with an identity for that message, and hence get the current key-ID.  Similarly, if a message is downloaded and marked as ``received'' using the plaintext tagging described earlier, the message disappears from the system, and this could be statistically associated with possible recipients.  This attack is possible even if new keys are regularly exchanged.  To mitigate these central points of attack, a full warp2 deployment ideally would have a distributed inbox database, and distributed storage with a suitable redundancy of messages and attachments to keep the memory footprint manageable.  This was not possible during the timeframe of the project, but potentially a desirable feature for future use.

\subsection{Vandalism \label{sec:vandalism}}
The prototype server accepts uploads from anybody, at the time of writing.  A malicious party who wishes to bring the server down could simply flood the server with messages to non-existent users, which nonetheless appear valid to the simplistic server code.  The inbox would grow indefinitely under this scenario.  In future production-ready deployments, a method of avoiding vandalism is required.

\subsection{Commercial Aspects}
At present, the strength of the method is in the fact that large numbers of people could use a warp2 server without identifying themselves.  Requiring login credentials weakens the method, because it creates unique identifiers before access to the inbox is allowed.  This also affects the commercialisation of the method, since paying customers would not want to be identified.

Instead of a closed system with login credential handling, warp2 is envisaged more along the lines of a TOR network.  Community-minded people would run warp2 servers that are open to the public.  These would be somewhat self regulating, since an over-burdened server with slow access and large numbers of messages would encourage migration of users to servers with lower loads.

Of course, a military deployment or commerical deployment within individual companies could run their own warp2 server very quickly, and restrict access by using ssh keys to limit logins to authorised client software only.  In this scenario, it is perhaps not such an issue to identify yourself as an employee by accessing the company warp2 server, just as a regular email system.  This is the foreseen deployment for journalists and governments.

\subsection{About the Prototype Application}
The prototype is written in C++ and Qt (Qt used under the LGPL).  The source code for the prototype is published on github \cite{WARP2_GITHUB} under the FreeBSD license.

The current status for platform support is as follows: the client code was developed on linux (fedora 20), and OS-X (mountain lion); the server code is extremely simple, running under apache and php5, and has been developed on OpenBSD and FreeBSD.

\section{Conclusions}
We have described and demonstrated a simple method of sending messages with encrypted addressing and message headers, protecting communications against interception and hindering the construction of network models of participants.  Weaknesses in the prototype implementation have been identified, and some mitigation strategies have been proposed for future work.

%
%
%
%
%
%
%

\section{Acknowledgements}
We are indebted to B. Hj\"{o}rvarsson and O. Kirstein for support in this project.

\bibliographystyle{elsarticle-num}

\end{document}